\documentclass[]{aa}
\usepackage{booktabs}
\usepackage{graphicx}
\usepackage{siunitx}
\usepackage{txfonts}
\usepackage{natbib}
\usepackage{amsmath}
\usepackage{color}
\usepackage{longtable}
\usepackage{lscape}
\usepackage{enumerate}
\usepackage{mathtools}
\usepackage{multirow}
\usepackage{multicol}
\usepackage{float}
\usepackage[version=4]{mhchem}
\newcommand\reaction[1]{\begin{align}\ce{#1}\end{align}}
\newcommand\reactionnonumber[1]%
 {\begin{align*}\ce{#1}\end{align*}}
\begin{document}
%-----------------------------------------------------------------------
% TITLE & AUTHORS
%----------------------------------------------------------------------
   \title{Destruction of dimethyl ether and methyl formate by collisions with He$^+$}
   \subtitle{} 
   \author{Daniela Ascenzi\inst{1}
     \and Andrea Cernuto\inst{1}
     \and Nadia Balucani\inst{2,3,4}
     \and Paolo Tosi\inst{1}
     \and Cecilia Ceccarelli\inst{3,4}
     \and Luca Matteo Martini\inst{1}
     \and Fernando Pirani\inst{2}}

   \offprints{D. Ascenzi, \email{daniela.ascenzi@unitn.it} N. Balucani, \email{nadia.balucani@unipg.it} C. Ceccarelli, \email{cecilia.ceccarelli@univ-grenoble-alpes.fr}}
       
   \institute{
     % 1
     Department of Physics, University of Trento, Via
     Sommarive 14, I-38123 Povo, Italy
     % 2
     \and Dipartimento di Chimica, Biologia e Biotecnologie,
     Universit\`a di Perugia, Via Elce di Sotto 8, I-06123 Perugia,
     Italy 
     % 3
     \and Univ. Grenoble Alpes, CNRS, IPAG, F-38000 Grenoble, France
     % 4
     \and  INAF-Osservatorio Astrofisico di Arcetri, Largo E. Fermi 5, I-50125, Florence, Italy
   }
   
   \date{Received November 2018; accepted March 2019}

   \titlerunning{DME and MF destruction by He$^+$}
   \authorrunning{Ascenzi et al.}

% \abstract{}{}{}{}{} 
% 5 {} token are mandatory
%=============================================
  \abstract
  % context heading (optional)
   {To correctly model the abundances of interstellar complex organic molecules (iCOMs) in different environments, both formation and destruction routes should be appropriately accounted for. While several scenarios have been explored for the formation of iCOMs via grain and gas-phase processes, much less work has been devoted to understanding the relevant destruction pathways, with special reference to (dissociative) charge exchange or proton transfer reactions with abundant atomic and molecular ions such as \ce{He+}, \ce{H3+} and \ce{HCO+}.}
  % aims heading (mandatory)
   {By using a combined experimental and theoretical methodology we provide new values for the rate coefficients and branching ratios (BRs) of the reactions of \ce{He+} ions with two important iCOMs, namely dimethyl ether (DME) and methyl formate (MF). We also review the destruction routes of DME and MF by other two abundant ions, namely \ce{H3+} and \ce{HCO+}.}
  % methods heading (mandatory)
   {Based on our recent laboratory measurements of cross sections and BRs for the DME/MF + \ce{He+} reactions over a wide collision energy range (Cernuto et al. 2017, Phys. Chem. Chem. Phys. 19, 19554; 2018 ChemPhysChem 19, 51), we extend our  theoretical insights on the selectivity of the microscopic dynamics to calculate the rate coefficients $k(T)$ in the temperature range from 10 to 298 K. We implement these new and revised kinetic data in a general model of cold and warm gas, simulating environments where DME and MF have been detected.}
  % results heading (mandatory)
   {Due to stereodynamical effects present at low collision energies, the rate coefficients, BRs and temperature dependences here proposed differ substantially from those reported in  KIDA and UDfA, two of the most widely used astrochemical databases. These revised rates impact the predicted abundances of DME and MF, with variations up to 40\% in cold gases and physical conditions similar to those present in prestellar cores.}
  % conclusions heading (optional), leave it empty if necessary 
   {This work demonstrates that the accuracy of astrochemical models can be improved by a thorough characterization of the destruction routes of iCOMs. The details of the chemical systems, indeed, can strongly affect their efficiency and significant deviations with respect to the commonly used Langevin model estimates are possible.}
% ----------------------------------
% --- Keywords  6 max---
% ----------------------------------

   \maketitle
%
%==============================================
\section{Introduction}\label{sec:introduction}

Interstellar complex organic molecules, hereafter called iCOMs, are C-bearing compounds containing at least six atoms  (\citealt{coms2009, ceccarelli2017}). Despite their relative simplicity with respect to what is considered a complex molecule in terrestrial terms, their detection in star forming regions (\textit{e.g.} \citealt{Rubin1971}; \citealt{blake1987}) has represented a challenge in our understanding of interstellar chemistry for decades. In particular, the attention has so far focused on the
possible formation routes, either via reactions occurring on the icy
surfaces of the interstellar grains (\textit{e.g.}
\citealt{garrod2006, ruuad2015, vasyunin2017}) or in the gas phase 
(\textit{e.g.} \citealt{RN7, RN1, skouteris2017, skouteris2018}).

Much less work has so far been spent on understanding the destruction pathways for these iCOMs, which also affect their abundances. Among all the possible destruction routes, three are not only efficient but often the dominant ones for all species, namely the reactions with He$^+$ and the two most abundant molecular ions \footnote{In water-rich regions, the H$_3$O$^+$ ion can also play a dominant role.}, H$_3^+$ and HCO$^+$.  He$^+$, H$^+$ and H$_2^+$ are created by the interaction of cosmic rays, which permeate the Galaxy, with H or He atoms, as well as H$_2$ molecules.  
The H$^+$ and H$_2^+$ ions formed in this way go on reacting with H$_2$ to produce H$_3^+$. The latter then reacts with CO (the most abundant molecule after H$_2$ in cold interstellar gas) and produces HCO$^+$.

Although the ionization cross sections of cosmic rays are  very small (\textit{e.g.} \citealt{padovani2009}), they are the dominant source of gas ionization in dense and UV-shielded environments. Likewise, despite their relatively low abundances ($\leq 10^{-6}$ with respect to H$_2$), He$^+$, H$_3^+$ and HCO$^+$ dominate the destruction of the vast majority of iCOMs. In general,
the reactions of iCOMs with H$_3^+$ and HCO$^+$ result in a proton
transfer which can be, but not necessarily is, dissociative. Protonated species, in turn, can undergo dissociative electron recombination, thereby producing neutral molecules.  In this last process, besides losing a H atom, the molecular species can undergo severe fragmentation (\textit{e.g.} \citealt{geppert2006}). Conversely, He$^+$ or H$^+$ react mostly by charge exchange. Because of the
large energy involved in the case of iCOMs, charge exchange is often
dissociative, with fragmentation likely more pronounced with He$^+$
due to the larger exotermicity. These processes are believed to be highly efficient and are included in all astrochemical models.

In spite of the importance of these destruction pathways of  iCOMs, studies devoted to characterizing the reaction rates are scarce. Of the almost one thousand reactions listed in the databases used in astrochemical models, KIDA\footnote{{\it http://kida.obs.u-bordeaux1.fr}} (KInetic Database for Astrochemistry: \citealt{RN53}) and UDfA\footnote{{\it  http://udfa.ajmarkwick.net/}} (UMIST Database for Astrochemistry: \citealt{mcelroy2013}), a rapid survey of how many of them have been
studied in detail, either in laboratories or theoretically, provides a
rather small number: no more than $\sim10$\%. For the vast majority of
the reactions with He$^+$, H$_3^+$ and HCO$^+$ both the rate
coefficients and product branching ratios (BRs) are guessed based on
common sense and analogous reactions. Very likely many of them are roughly correct, but it is difficult, if not impossible, a priori to say so.

\begin{figure*}
\centering
\includegraphics[width=14cm,angle=0]{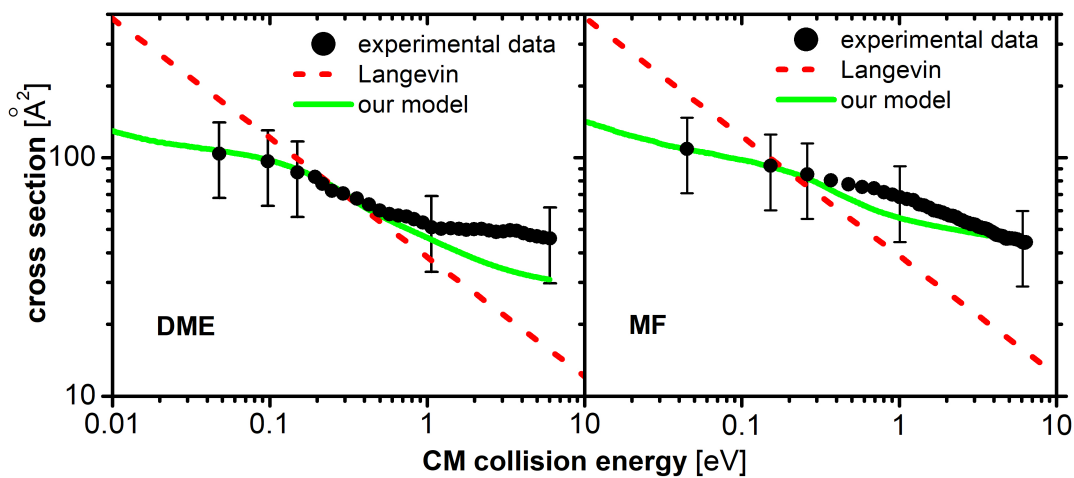} 
\caption{Experimental and theoretical cross-sections for the charge-exchange reaction of \ce{He+}-DME ({\it left panel}) and \ce{He+}-MF ({\it right panel}), as a function of the collision energy. The black
  points represent the measured total cross-sections
  \citep{C7CP00827A, cernuto2018}. Error bars represent a 35\% error 
  on the absolute values of the cross-section (relative cross-sections
  have smaller uncertainties). The curves represent the total
  cross-sections calculated by our model, which is based on an
  improved Landau-Zener-St\"{u}ckelberg approach (green lines), and a
  simple Langevin model (red dashed lines), (see text for
  details).}
 \label{fig:1}
\end{figure*}

With this work, we want to start filling this crucial gap in our knowledge of astrochemistry. Specifically, we report an experimental and theoretical study of the reaction of He$^+$ with two important iCOMs: dimethyl ether (CH$_3$OCH$_3$: hereinafter DME) and methyl formate (HCOOCH$_3$: hereinafter MF). We chose these two species for the following reasons:
\begin{enumerate}
\item They are detected in hot cores/corinos and even cold prestellar
  cores with abundances, with respect to \ce{H2}, that range from
  $\sim 10^{-10}$ to $\sim10^{-7}$ (\citealt{jaber2014, jimenez2016}).
\item Two competing theories postulate that DME and MF are formed
  either on the grain surfaces (\textit{e.g.} \citealt{garrod2006,shingledecker2018}) or in the gas phase (\citealt{RN1}, \citealt{Skouteris2019}), so that their study can give us a guidance as to which theory is more correct.
  
\item No measurements or theoretical studies exist in the literature
  about the DME and MF destruction by He$^+$.
\item Previous works have provided information on the rate
  coefficients and product BRs of the reactions of DME and MF with
  H$_3^+$ and HCO$^+$ (\textit{e.g.} \citealt{RN12, lee1992,
    lawson2012}), as well as on the dissociative recombination with electrons of methoxymethyl and protonated DME cations (\textit{e.g.} \citealt{hamberg2010}), so that a comparison is possible.

\item They are targets of the large observational program IRAM-NOEMA SOLIS\footnote{{\it https://solis.osug.fr/}} (\citealt{ceccarelli2017}), so that their abundances are and will be measured in several different environments, which will allow us to test the various formation theories (point 1), {\it if, and only if, we can be sure of the destruction routes}.
\end{enumerate}
The present article has been stimulated by recent measurements
performed by some of the present authors under high resolution
conditions of the absolute total cross-section for the reaction of
\ce{He+} with DME and MF in a wide range of collision energies
 \citep{C7CP00827A, cernuto2018}. The results of this study are
summarized in Fig.~\ref{fig:1} and briefly described in Section
\ref{sec:methodology}. Detailed analysis of these results suggests that the dynamical evolution of such systems is driven, at a microscopic level, by a strong anisotropy in the interaction potential energy hypersurfaces, which presumably also affects their macroscopic behaviour.

In this article, we first review the rate coefficient and BRs values for the DME + He$^+$ and MF + He$^+$ reactions that are available in the astrochemical databases KIDA and UDfA (Sec. \ref{sec:brief-revi-exist}). We then provide new rate coefficients and BRs, based on our recent laboratory measurements and theoretical insights on the selectivity of the microscopic dynamics (Sec. \ref{sec:new-reaction-rates}). For completeness, we also review
the reactions of DME and MF with the other molecular ions abundant in
the interstellar gas, namely HCO$^+$ and H$_3^+$
(Sec. \ref{sec:revis-valu-react}). Finally, we discuss the
implications for astrochemical model predictions
(Sec. \ref{sec:astr-impl}) and the major conclusions of this study
(Sec. \ref{sec:conclusions}).

%===============================================
\section{Brief review of the DME and MF reactions with He$^+$ in the
  astrochemical databases}\label{sec:brief-revi-exist}

\subsection{Methodology}\label{sec:gener-cons-how}
In the KIDA and UDfA databases, the suggested
values for rate constants as a function of temperature are based on a
modified Arrhenius equation, using the parameters from the OSU2009
 gas-phase chemistry database of Herbst \citep{RN55}.
\begin{align}
k(T)=\alpha \left( \frac{T}{300} \right) ^{\beta} e^{- \gamma / T} \; \;[\si{\cubic\centi\meter\per\second}]
\label{eq:1}
\end{align}
For ion-neutral reactions, $\alpha$ is a parameter specific for each
reaction, while the other two parameters are usually assumed to be
$\beta$=$-0.50$, $\gamma$=$0.00$ and the temperature $T$ is in the range from \SIrange{10}{280}{K} (in the KIDA database) or from  \SIrange{10}{41000}{K} (in the UDfA database).  

It should be noted that the above expression is
an approximation (valid for low temperatures) of a more general
expression deriving from a classical dynamic treatment of capture in a
non central potential.  Such a capture model has been developed by Su
and Chesnavich \citep{su1980, su1982} and it is a combined
variational transition state theory/classical trajectory study of
thermal energy collisions between an ion (treated as a point charge
particle) and a polar molecule (treated as a two-dimensional rigid
rotor).  In the model, the interaction potential between the ion and
the polar molecules includes only the ion-induced dipole and
ion-permanent dipole terms.

Thermal rate coefficients determined from the calculations have been fitted to analytical expressions given in terms of the parameter $x$ defined as:
\begin{align}
x= \mu_D / (2\alpha_e k_B T)^{0.5}	
\end{align}
where $\mu_D$ is the dipole moment of the neutral, $\alpha_e$ is the
average electronic polarizability of the neutral and $k_B$ is the
Boltzmann constant.  

For $x\geq2$, the Su \& Chesnavich formula for
the rate constant $k(T)$ can be written in the following form:
\begin{align}
k(T) = c_1 + c_2T^{-0.5}
\label{eq:2}
\end{align}
with $c_1=0.62k_L$ and $c_2 = 2.1179 (\mu_D q)/(\mu k_B)^{-0.5}$ and where $k_L$ is the Langevin rate constant:
\begin{align}
k_L = 2\pi q(\alpha_e/\mu)^{0.5}
\label{eq:lang}
\end{align}
with $\mu$ being the reduced mass of the reactants and $q$ the
electric charge of the ion.  In expression~(\ref{eq:2}), when cgs-esu
units are utilised, the $k(T)$ values are obtained in
\si{cm^3~s^{-1}}. Additionally, when $c_1\ll c_2 T^{-0.50}$, which is more likely at low temperatures, equation~(\ref{eq:2}) reduces
to~(\ref{eq:1}).  

For $x < 2$ a more complex expression than Eq.~(\ref{eq:2}) is
obtained (see \citealt{su1982} and \citealt{Wakelam2010} for the detailed 
expressions).

The Su \& Chesnavich model has been reported
to work well for neutral species with linear, symmetric tops or
asymmetric top geometries. However, being a classical treatment, it is valid
only at temperatures above which the rotational motion can be
considered classical, \textit{i.e.} when $k_BT$ is much higher than
the relevant rotational constants of the neutral molecule.  For most
species, the classical treatment is reliable down to 10-20 K, while at
lower temperatures ($\sim$ 0.1-10 K) a semi-classical regime applies
in which quantum effects in the hindered-rotor type motion should be
considered, while the relative translational motion can still be
treated classically. Using the statistical adiabatic channel model
developed by J. Troe and collaborators \citep{troe1996}, the capture of rotationally state-selected and unselected polar molecules by ions
in such low temperature regimes has been studied in details
(\citealt{Maergoiz2009} and references therein).  The noticeable result 
is that an expression similar to Eq.~(\ref{eq:2}) can be used where
$\mu_D$ is substituted by an effective dipole moment of the neutral
reactant ($\mu_{D,eff}$) which needs to be calculated following the
lines indicated in \cite{Maergoiz2009}.  However, the differences
might be so small (in the case of \ce{H2O} as neutral partner
$\mu_{D,eff}\approx 1.07 \mu_D$) that they lie within the uncertainty
of the fits and calculations using the Su \& Chesnavich model.

\subsection{The case of dimethyl ether}

In the KIDA and UDfA databases, the reaction of He$^+$ ions with DME
is reported to give the following two products, each with a BR of
50.0\%:
\reaction{CH3OCH3 + He+ &-> H + He + H2CO + CH3+ \label{reac:1}\\
  CH3OCH3 + He+ &-> He + CH3 + H2COH+ \label{reac:2} } 
The suggested values for the total rate constant as a function of
temperature are given in terms of the modified Arrhenius equation
(\ref{eq:1}), with the following parameters:
$\alpha=2.64\times 10^{-9}$ in KIDA and $\alpha=2.00\times 10^{-9}$ in
UDfA, with $\beta = -0.50$, $\gamma = 0.00$ in both.  Also in this case, the $k(T)$ expression reported in the KIDA and UDfA databases is
an approximation (valid for low temperatures) of the more general
equation~(\ref{eq:2}).

In the DME case ($\mu_D=1.3$ D, $\alpha_e$=5.16 \AA$^3$,
$\mu=6.112\times 10^{-27}$ kg), the Langevin rate constant is
$k_L= 2.77\times 10^{-9}$ cm$^3$s$^{-1}$ and the use of
expression~(\ref{eq:1}) rather than (\ref{eq:2}) at $T=10$ K
leads to an underestimation of the rate constant by about 10\%, a value which rises to about 39\% at $T=280$ K.

\begin{figure*}[!ht]
\centering
\includegraphics[angle=0,width=12cm, angle=0]{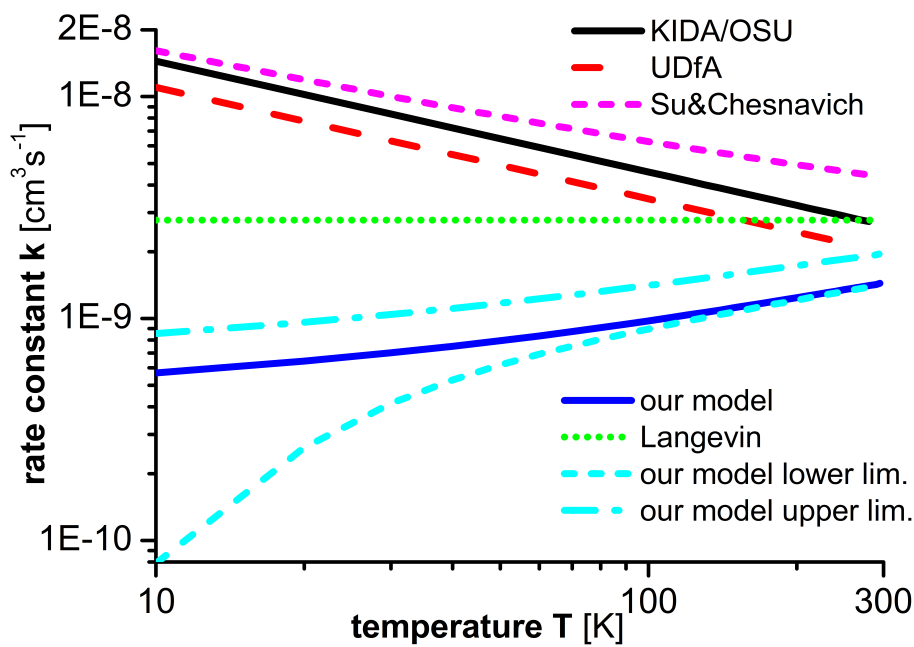} 
 \caption{Temperature dependence of the total rate constant $k$ for
   the reaction of He$^+$ with DME. Solid black line: recommended values from KIDA database according to Eq.~(\ref{eq:1}); dashed red line: recommended values from UDfA according to
   Eq.~(\ref{eq:1}); dotted green line: Langevin rate constant
   $k_L$ according to Eq.~(\ref{eq:lang}); dashed pink line: classical trajectory scaling according to Eq.~(\ref{eq:2}) \citep{su1982}; solid blue line: our calculations; cyan dotted-dashed line: our calculations, upper limit estimate; cyan dashed line: our calculations, lower limit
   estimate (see text).}
 \label{fig:2}
\end{figure*}

The rate constants as a function of temperature over the 10-300 K range are calculated according to the recommendation of KIDA and UDfA databases, as well as using the trajectory scaling method of
Eq.~(\ref{eq:2}), and the values are reported in Fig.~\ref{fig:2}.  

Finally, the KIDA database gives
$k$=\SI{7.23e-9}{\cubic\centi\meter\per\second} for both
reactions~(\ref{reac:1}) and (\ref{reac:2}) at \SI{10}{K}, with a total
$k$ for the title reaction equal to
\SI{1.45e-8}{\cubic\centi\meter\per\second}.  At \SI{280}{K} (upper
limit of Eq.~(\ref{eq:1})), the model gives a total $k$ of
\SI{2.74e-9}{\cubic\centi\meter\per\second}.  In the UDfA database,
the $k_{\SI{300}{K}}$ is fixed to
\SI{2.00e-9}{\cubic\centi\meter\per\second}, giving a rate constant at
\SI{10}{K} equal to \SI{1.10e-8}{\cubic\centi\meter\per\second} and
\SI{2.07e-9}{\cubic\centi\meter\per\second} at \SI{280}{K}.

\subsection{The case of methyl formate}

In the KIDA and UDfA databases, the reaction of He$^+$ ions with MF is
reported to give only the following channel (hence with a BR of 100.00
\%):
\reaction{HCOOCH$_3$ + He$^+ ~\rightarrow$ He + CH$_3$ +
  HOCO$^+$ \label{reac:3}} 
The suggested values for the rate constant as a function of
temperature are given in terms of the modified Arrhenius equation
(\ref{eq:1}), with the following parameters:
$\alpha=3.54\times 10^{-9}$ in KIDA and $\alpha=3.00\times 10^{-9}$ in
UDfA, with $\beta = -0.50$, $\gamma = 0.00$ in both.  Also in this
case, the $k(T)$ expression reported in the KIDA and UDfA databases is
an approximation (valid for low temperatures) of the more general
equation~(\ref{eq:2}).

For MF, ($\mu_D=1.77$ D, $\alpha_e=5.3$ \AA$^3$,
$\mu=6.227\times10^{-27}$ kg) the Langevin rate constant is
$k_L= 2.78\times10^{-9}$ cm$^3$s$^{-1}$ and the use of
expression~(\ref{eq:1}) rather than (\ref{eq:2}) at $T=10$ K
leads to an underestimation of the rate constant by about 8\%, a value which rises to about 32\% at $T=280$ K.

\begin{figure*}[!ht]
\centering
 \includegraphics[angle=0,width=12cm,angle=0]{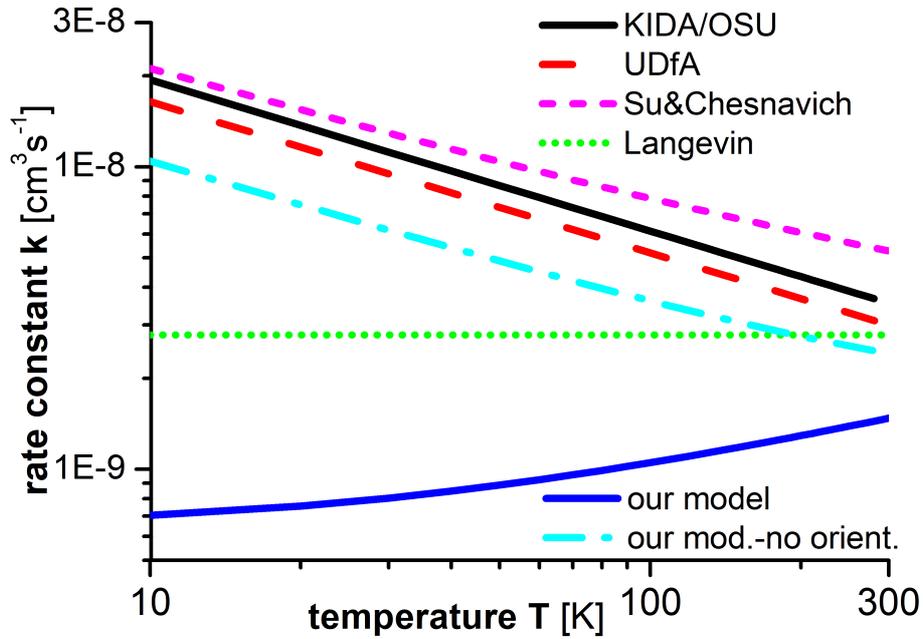} 
 \caption{Temperature dependence of the total rate constant $k$ for
   the reaction of He$^+$ with MF. The curve colors are as in
   Fig.~\ref{fig:2}, except for the dashed-dotted cyan line which corresponds to the data calculated using our model and assuming no preferential orientation of the colliding system.}
 \label{fig:3}
\end{figure*}

Rate constant values as a function of temperature in the range 10-300 K were calculated using the expressions reported in the KIDA and UDfA databases, as well as with the trajectory scaling method of Eq.~(\ref{eq:2}) and results are shown in Fig.~\ref{fig:3}.

Finally, the KIDA and UDfA values are equal to
\SI{1.94e-8}{\cubic\centi\meter\per\second} and
\SI{1.64e-8}{\cubic\centi\meter\per\second}, respectively, at
\SI{10}{K}, and \SI{3.66e-9}{\cubic\centi\meter\per\second}
and \SI{3.11e-9}{\cubic\centi\meter\per\second} at \SI{280}{K}.

%===============================================
\section{New reaction rate coefficients and branching ratios for \ce{He+}/DME and \ce{He+}/MF}\label{sec:new-reaction-rates}

\subsection{Our model of the total cross-sections}\label{sec:methodology}

As already mentioned in the Introduction, absolute total
cross-sections as a function of the collision energy in the 0.04-6~eV
energy range for collisions of He$^+$ ions with DME and MF were
recently measured using a guided ion beam mass spectrometer
(\citealt{C7CP00827A, cernuto2018}; see Fig. \ref{fig:1}). 
In order to both analyze the results and properly rationalize all our experimental findings, this work focuses on two main aspects.

First, the entrance and exit channels of the potential energy surfaces
(PESs) of the two reactions were characterized.  The non-covalent
nature of the interactions implies that each PES will depend on the balance between repulsion and the combined dispersion and induction attraction.
For the entrance channels, the electrostatic
contribution due to the long range ion-permanent dipole interaction
components has also been added. We found that the interaction
anisotropy becomes comparable to or larger than the mean rotational
energy of DME and MF at room temperature, already at large
separation distance from \ce{He+} (\textit{i.e.} 15-20 \AA).
Therefore, the approach of \ce{He+} during the collision leads each
molecule to probe an increasing electric field gradient that becomes
sufficient to transform the molecular free rotation into hindered
rotation and, afterwards, to drive the system in a pendular state.
The latter is a particular case of bending motion confined in narrow angular cones around the most stable configuration of the ion-molecule collision complex. Such "natural" molecular polarization phenomena become important in systems driven by strongly anisotropic long-range forces, as is the case for the present reactions. These effects are expected to grow in significance at low collision energies, causing changes in reactivity due to changes in the structure and stability of the collision complexes. They can therefore be regarded as "stereodynamical effects".
%they can favour or hinder reactivity, since they are controlled by the structure and stability assumed by the collision complex. For this reason they can also be referred to as "stereodynamical effects".

Secondly, we used a semi-classical model to evaluate the transition
probability from the entrance to the exit channels. The model was
developed \textit{ad hoc} to give a unified description of the collision dynamics of both systems and to interpret the experimental
results shown in Fig.~\ref{fig:1}.  In the model, the transfer of an
electron from DME/MF to He$^+$ is assumed to proceed via non-adiabatic transitions located at the crossings between the entrance
(\ce{He+}-DME/MF) and exit (He-\ce{DME^{.+}}/\ce{MF^{.+}}) sections of the PESs given in the diabatic representation.  Positions and energies
of the crossings indicate that \ce{He+} captures an electron exclusively from an inner molecular orbital of DME/MF, thus generating a highly excited/unstable molecular ion that can readily dissociate.  

The probability of the non-adiabatic event at each crossing has been evaluated using a generalized Landau-Zener-St\"{u}ckelberg approach, taking into account the symmetry of the electron density distribution of the molecular orbitals of DME/MF from which the electron is removed. To this end, the non-adiabatic coupling has been represented, according to previous treatments \citep{gislason1975,gislason2007, falcinelli2016}),
as a constant value modulated by an angular term. The latter brings the coupling to zero for geometries with null overlap between the atomic \textit{s} orbital of He$^+$ and the relevant molecular orbitals, while it makes the coupling maximum for geometries giving the best overlap (corresponding to the maximum electron density for the relevant molecular orbitals).
In some cases, it was necessary to assume that, in absence of an appreciable overlap between the molecular and atomic orbitals exchanging the electron, the non-adiabatic
events are exclusively due to the Coriolis (rotation-orbit)
coupling. 

As demonstrated by the data plotted in Fig. \ref{fig:1}, our approach is able to reproduce the experimental cross sections (especially at low collision energies), while a simple Langevin model fails.

\subsection{Rate coefficients}
\subsubsection{Computational method}
The rate coefficients as a function of temperature, $k(T)$, have been obtained by averaging the total cross-sections calculated using our model (briefly described in Section \ref{sec:methodology} and shown in Fig. \ref{fig:1}) over a Maxwell-Boltzmann distribution of collision energies, $E$, between $E_{min}$ and $E_{max}$, as follows:
\begin{align}
\centering
k(T) = \left( \frac{1}{\pi \mu} \right) ^{1/2} \cdot \left( \frac{2}{k_{B} T} \right) ^{3/2} \cdot \int_{E_{min}}^{E_{max}}{\sigma(E) \cdot E \cdot e^{- E /k_{B} T } dE}
\label{MBdistr}
\end{align}

This treatment, based on a generalized Landau-Zener-St\"{u}ckelberg
approach, defines the energy range within which the cross-sections
calculated using a semi-classical method are valid. Specifically,
the adopted semi-classical methodology is applicable
when the \textit{de Broglie} wavelength ($\lambda _{DB}$) for the
relative motion of the colliding partners is smaller than the
distances at which the potential wells in the entrance and exit
channels are located.  For example, for DME these distances range
between \SIrange{1.5}{3}{\angstrom}, while for MF they range between \SIrange{2}{4}{\angstrom}. Crossings among the entrance and exit potential energy curves are located within the same range. To obtain a conservative estimate, we have assumed that $\lambda _{DB}$=\SI{1.5}{\angstrom}, resulting in a minimum energy $E$ of about \SI{10}{meV}, below which the semi-classical treatment is not fully reliable.

It is worth noting that, to obtain an accurate estimate of $k(T)$ at low temperatures (without incurring errors from numerical artefacts), cross-section values at energies smaller than \SI{10}{meV} have to be included in the integral of Eq. \ref{MBdistr}.  

\subsection{Dimethyl ether}
In order to evaluate the most appropriate $E_{min}$ to use in the integral of
Eq. (\ref{MBdistr}), we compared the result of Eq. (\ref{MBdistr})
when $\sigma(E)$ is taken to be equal to the Langevin cross-section
$\sigma_{Langevin} (E)=q \sqrt{\frac{\pi \alpha}{2 \varepsilon_0 E}}$
(shown in Fig. \ref{fig:1}) with the rate coefficient given by the
Langevin model (Eq.~(\ref{eq:lang})).

We carried out these computations for the DME case, for which the
Langevin rate coefficient is \SI{2.77e-9}{\cubic\centi\meter\per\second}, and it is independent of $T$.
We fixed the upper energy limit $E_{max}$ at \SI{10000}{meV} and
varied $E_{min}$ to obtain an estimate of the error. At 300 K, the numerical error on the calculated rate coefficient is negligible already when $E_{min}$ is set equal to \SI{0.1}{meV}. However, at 10 K, it is necessary to decrease $E_{min}$ to \SI{0.01}{meV} to have a relative error within 0.1\%.

Therefore, we set $E_{max}$ to \SI{10000}{meV} and $E_{min}$ to \SI{0.01}{meV} in order to compute the rate coefficients with the cross-sections from our model (\citealt{C7CP00827A, cernuto2018}: see Section
\ref{sec:methodology}). However, since our cross-sections would strictly
apply only up to energies larger than about 10-30 meV, we also calculated the rate coefficients over the 0.01 to 10 meV range using two different models for the 
cross-section over this range. Such models are described below:
%computed the rate coefficients assuming two additional (extreme) behaviours of the cross-section in the 0.01 to 10 meV range:
%
\begin{enumerate}
\item Cross-sections increasing with increasing energy $E$: a function
  $\varpropto E^2$ has been chosen to represent $\sigma$;
\item Cross-sections decreasing with increasing  energy $E$: in this case, the Landau-Zener-St\"{u}ckelberg treatment has been used, but the strength of the Coriolis coupling has been modulated by obtaining calculated cross-sections at the upper limit of the experimental determinations.
\end{enumerate}
Using the cross-sections from the first case ($\sigma \varpropto E^2$),
the resulting rate coefficient at \SI{10}{K} is lower by a factor of 7.7
with respect to the one calculated with our model, whereas at
\SI{300}{K} the two values coincide. Using the cross-sections from the
second case, the resulting rate coefficient is higher by a factor
1.4--1.5 over the whole range of temperature.  The resulting curves are
shown in Fig.~\ref{fig:2}: the first case likely represents a lower limit
while the second case is likely an upper limit to the rate
coefficients, as labelled in the figure. Similar values for upper and lower limits to $k$ are expected to be valid also for the MF case (see next Section).

\subsubsection{Methyl formate}
%CC: The following sentence does not mean much to me, so I commented it
%Simulations for MF have not carried out, but we expect that the rate
%coefficients can vary also in this case in the same range of values.

In the MF case, in addition to computing the rate coefficients using our model for the cross-sections (Section \ref{sec:methodology}), we have also performed a simulation without the assumption of a strong stereochemical effect in the reaction induced by the formation of pendular states (\textit{i.e.} preferred orientations of the polar molecule in the electric field generated by He$^+$), and neglecting the Coriolis coupling. Should this be the case, the dynamical calculations
must include contributions to the cross-sections from all the
possible approach geometries of He$^+$ and MF.
Therefore, contributions from points over the whole sphere around the neutral species must be considered rather than just selected contributions from a limited number of configurations within narrow angular cones around the most attractive geometries, such as is the case for our
model.  For a proper mapping of all the possible approach geometries,
a total of 74 equally spaced points on the sphere around the center of mass of the reacting system have been considered.

The resulting rate coefficients obtained with the cross-sections calculated independent of orientation are shown in Fig. \ref{fig:3}, with the label ``our mod.-no orient.''. The
rate coefficient at 10 K is more than one order of magnitude larger than the one computed using our model that considers the orientation dependence of reactivity. The difference between the two estimates diminishes with increasing temperature but it is still about 30\% at 300 K.

We emphasize, however, that the cross-sections calculated assuming an
important orientation effect among the reactants and the formation of
pendular states, i.e. the model of Section \ref{sec:methodology}, are
in good agreement with the experimental results (Fig. \ref{fig:1}).
Conversely, calculations in the absence of stereo-chemical effect
are not able to reproduce the experimental results. This provides a strong
evidence that stereo-chemical effects play a role in the collisions of
He$^+$ with polar neutrals and should not be disregarded when seeking to obtain reasonable estimates of rate constants. In the following, therefore,
we will use the rate coefficients obtained using our model from Section \ref{sec:methodology}.

\subsection{Comparison with previous rate coefficients}

Our new rate coefficients are compared with recommended
values from KIDA, UDfA and classical trajectory scaling for the
He$^+$-DME case in Fig. \ref{fig:2}, while the corresponding values
for the He$^+$-MF case are presented in Fig. \ref{fig:3}.
The differences between the calculated rate coefficients and the
estimates of KIDA and UDfA databases are evident.  

For DME, we obtain values of $k$=\SI{5.69e-10}{\cubic\centi\meter\per\second} at
\SI{10}{K} and
%$k$=\SI{1.41e-9}{\cubic\centi\meter\per\second} at \SI{280}{K}
\SI{1.44e-9}{\cubic\centi\meter\per\second} at \SI{300}{K}.  At
\SI{10}{K}, the KIDA database overestimates the rate coefficient by a
factor $\sim$ 26, while the UDfA value is $\sim$ 20 times higher than our rate. At \SI{300}{K}, the differences  between the calculated and estimated rate constants decrease, with the KIDA value being $\sim$ 2 times higher than our value and the UDfA one being only $\sim$ 1.5 higher.

For MF, we obtain values of $k$=\SI{7.05e-10}{\cubic\centi\meter\per\second} at
\SI{10}{K} and 
%$k$=\SI{1.44e-9}{\cubic\centi\meter\per\second} at\SI{280}{K} 
\SI{1.48e-9}{\cubic\centi\meter\per\second} at
\SI{300}{K}.  At \SI{10}{K}, KIDA and UDfA recommend values are $\sim$ 27 and $\sim$ 23 times higher, respectively, than our value. Around
\SI{300}{K}, KIDA and UDfA overestimate the rate constants by factors of $\sim$ 2.5 and $\sim$ 2, respectively. 

In summary, our results show the database values to be significant overestimates, especially at \SI{10}{K}.
It is worth noting that the rate coefficients of the astrochemical
databases exhibit a negative temperature dependence
(\textit{anti}-Arrhenius behavior), while those calculated by us show
a positive Arrhenius dependence, even though the present charge
transfer processes are both exothermic and barrierless. This is at odds
with the common understanding of ion-molecule reactions, but our model provides a rationale for such apparent incongruity. The strong anisotropic interactions in the  He$^+$-DME/MF systems drive the collision complexes into the most attractive configurations, which happen to be the least efficient for charge exchange due to unfavourable molecular overlap. 
%\textcolor{red}{Seguendo il consiglio di Cecilia ho aggiunto la frase sopra per tranquillizzare il lettore sull'apparente misbehaviour. }
%CC: can we add a sentence to pacify the reader on why this apparent misbehaviour?

\subsection{Branching ratios}

The difference between our measured BRs \citep{C7CP00827A, cernuto2018}
 and those reported in the KIDA and UDfA databases is striking, as shown in the Tables~\ref{table1} and \ref{table2}. The uncertainties on our experimental BRs derive from error propagation in the cross section measurements.

 \begin{table}[!h]
\centering
 \begin{tabular}{lcccc}
\toprule
\multirow{2}*{Product} & \multirow{2}*{\textit{m/z}} & \multicolumn{3}{c}{BRs (\%)} \\
\cmidrule(lr){3-5}
& & Exp. Values & KIDA & UDfA \\
\midrule
\ce{CH2^{+.}}   & 14  & 7.3$\pm$0.8 & 0& 0\\
\ce{CH3+} & 15  & 38.5$\pm$3.7 & 50& 100\\
\ce{HCO+} & 29  & 53.6$\pm$5.3 & 0& 0\\
\ce{OCH3+} & 31 & 0.4$\pm$0.7  & 50& 0\\
\bottomrule
\end{tabular}
 \caption{Product branching ratios (BRs) for the reaction He$^+$ + DME.
 Values are obtained, at a collision energy in the CM frame of $\sim$
 1.6~eV, by averaging results measured at  pressures of DME between 2$\times 10^{-7}$ and 5$\times 10^{-6}$~mbar. BRs show no change with varying collision energy over the explored range.}
  \label{table1}
\end{table}

 \begin{table}[!h]
\centering
\begin{tabular}{lcccc}
\toprule
\multirow{2}*{Product} & \multirow{2}*{\textit{m/z}} & \multicolumn{3}{c}{BRs (\%)} \\
\cmidrule(lr){3-5}
& & Exp. Values & KIDA & UDfA \\
\midrule
\ce{CH2^{+.}}   & 14  & 3.6$\pm$0.3 & 0& 0\\
\ce{CH3+} & 15  & 7.3$\pm$0.3 & 0& 0\\
\ce{HCO+} & 29  & 83.2$\pm$2.0 & 0& 0\\
\ce{OCH3+} & 31 & 4.2$\pm$0.8  & 0& 0\\
\ce{CO2^{+.}} & 44 & 1.3$\pm$0.1  & 0& 0\\
\ce{HC(O)O+} & 45 & 0.4 $\pm$0.1  & 100& 100\\
\bottomrule
\end{tabular}
  \caption{Product branching ratios (BRs) for the reaction He$^+$ + MF.
Values are obtained, at a collision energy in the CM frame of $\sim$
0.9~eV, by averaging results measured at  pressures of MF between  1.8$\times 10^{-7}$ and  1.4$\times 10^{-6}$~mbar. BRs show no change with varying collision energy over the explored range.}
  \label{table2}
\end{table}

In the DME case, while the KIDA database assumes a BR of 50.0\% for
the production of CH$_3^+$ and OCH$_3^+$, our results suggest that
the most abundant product is HCO$^+$ (BR=53.6\%) and that CH$_3^+$ and
OCH$_3^+$ have branching ratios of 38.5\% and 0.4\%,
respectively. Results for MF show an even greater disagreement, with HCO$^+$
being the most abundant product according to our experiments (with a
BR=83.2\%), while both KIDA and UDfA assume the exclusive formation of
HOCO$^+$.

%%%%%%%%%%%%%%%%%%%%%%%%%%%%%%%%%%%%%%%%%%%%%%%%%%%%%%%%%%%
\section{Revised values for the reactions of \ce{HCO+} and \ce{H3+} with DME and MF}\label{sec:revis-valu-react}

For completeness, we review here the rate coefficients for the other two ions which dominate 
the destruction pathways for iCOMs (see Introduction).

\subsection{\ce{HCO+ + DME/MF}}

An experimental determination of the rate constants for the proton
transfer reaction of \ce{HCO+} with DME and MF at \SI{298}{K} exists
\citep{RN12} and gives the following results:
$k=(2.1 \pm 0.5) \times 10^{-9}$~cm$^3$s$^{-1}$ for DME and
$(2.9 \pm 0.7) \times 10^{-9}$~cm$^3$s$^{-1}$ for MF.  In both cases,
the proton transfer is non dissociative.  

However, the KIDA database recommends to use the value from the modified Arrhenius equation (see Eq.(\ref{eq:1})) with the following parameters: $\alpha$=$1.2 \times 10^{-9}$ for DME and $1.55 \times 10^{-9}$ for MF, with $\beta$=$-0.50$ and $\gamma$=$0.00$ for both. At 298~K the resulting rate constants are $k=1.2 \times 10^{-9}$~cm$^3$s$^{-1}$ for DME and $1.56 \times 10^{-9}$ ~cm$^3$s$^{-1}$ for MF, \textit{i.e.} smaller than the measured values by about a factor of two.

Conversely, the UDfA database suggests using the experimental data of Tanner \textit{et al.} at 298~K and rescales them with temperature according to Eq.(\ref{eq:1}), giving $k_{300}=2.1 \times 10^{-9}$~cm$^3$s$^{-1}$ (for DME) and $ 2.9 \times 10^{-9}$~cm$^3$s$^{-1}$ (for MF), with $\beta$=$-0.50$ for both. At 10~K the resulting rate constants are $1.15 \times 10^{-8}$~cm$^3$s$^{-1}$ for DME and $1.58 \times 10^{-8}$~cm$^3$s$^{-1}$ for MF. For both systems  accuracy in the rate constants is given within $\pm$ 25\%). In the following section, we will adopt the UDfA recommendations.

\subsection{\ce{H3+ + DME/MF}}

{\it Dimethyl ether:}\\
Rate coefficients and BRs for the reaction of \ce{H3+} with DME have been measured using a flowing afterglow technique at 300~K \citep{lee1992} and the results are as follows: the overall rate coefficient is $4.7 \times 10^{-9}$~cm$^3$s$^{-1}$ and the proton transfer is highly dissociative giving the following fragments, with \% BRs in brackets: \ce{CH3+} (29\%), \ce{CH5+} (8\%),
\ce{HCO+/C2H5+} (10\%), \ce{CH3O+} (26\%), \ce{C2H5O+} (15\%),
\ce{CH3OHCH3+} (12\%).  

The KIDA data is at odds with these findings, as it exclusively quotes protonated DME (\ce{CH3OHCH3+}) as a product with $k=3.01 \times 10^{-9}$~cm$^3$s$^{-1}$ at 298~K (from the modified Arrhenius equation (\ref{eq:1}) with parameters $\alpha$=$3.00 \times 10^{-9}$, $\beta$=$-0.50$, $\gamma$=$0.00$).
The UDfA database similarly gives a rate constant for non dissociative proton transfer in terms of Eq.(\ref{eq:1}) with $k_{300}=2 \times 10^{-9}$ and $\beta$=$-0.50$ for T ranging from 10 to 41000~K, with an accuracy of $\pm$50\% in the rate constant.

Hereafter, we will use the total $\alpha$=$4.70 \times
10^{-9}$, BRs as given by \cite{lee1992}, $\beta$=$-0.50$ and $\gamma$=$0.00$ over the 10--300 K temperature range. In order to avoid introducing new trace ions into the network, we will consider only the two most abundant channels, namely those giving CH$_3^+$ and CH$_3$O$^+$ as products, and the channel relative to the undissociative proton-trasfer leading to CH$_3$OHCH$_3^+$ plus H$_2$ (already included in the models with a different rate coefficient).

\noindent
{\it Methyl formate:}\\
In the MF case no measurements for the rate coefficients have been performed. However, Lawson \textit{et al.} \citep{lawson2012}, while  attempting to measure the rate constants for dissociative electron-ion recombination of protonated MF, observed that significant fragmentation occurred when attempting to induce proton transfer from \ce{H3+} to MF. In addition, according to a chemical ionization mass spectrometric experimental study \citep{Herman81}, the outcome of the protonation reaction of MF by H$_3^+$ is mostly dissociative and the observed relative abundances of ionized fragments are:
15.5\% for H$_5$C$_2$O$_2^+$, 74\% for CH$_3$OH$_2^+$, 5.9\% for CH$_2$OH$^+$ and 4.4\% for CH$_3^+$. 

Such experimental evidences are at odds with the prescriptions from KIDA, where the reaction is assumed to be completely non dissociative with $k=4.06 \times 10^{-9}$~cm$^3$s$^{-1}$ at 298~K (from the modified Arrhenius Eq.(\ref{eq:1}) with parameters $\alpha=4.05 \times 10^{-9}$, $\beta$=$-0.50$ and $\gamma$=$0.00$).  
The value reported in the UDfA database is Eq.(\ref{eq:1}) with $k_{300}=3 \times 10^{-9}$ and $\beta$=$-0.50$ (for a T range 10-41000~K).

%In absence of experimental or theoretical data, in the following, we
%will adopt the KIDA prescription.

In the following, we will adopt the KIDA value for the total rate constant ($i.e.$ $\alpha=4.05 \times 10^{-9}$, $\beta$=$-0.50$ and $\gamma$=$0.00$), while for the BRs we will use the values found by \cite{Herman81}. As before, to avoid introducing new trace ions in the network, we will consider only the two most abundant channels, namely those producing CH$_3$OH$_2^+$ and protonated MF.

%===============================================
\section{Astrophysical implications}\label{sec:astr-impl}

\subsection{New values to adopt in the astrochemical networks}

From the discussion in \S \ref{sec:new-reaction-rates}, it is clear
that the new rate coefficients, based on the experimental results by
\cite{C7CP00827A, cernuto2018}, differ from the those reported in 
both KIDA and UDfA databases in terms of their absolute values, their
temperature dependence (which is positive rather than negative) as
well as in the product BRs.

An interpolation of the new rate coefficients (and extrapolated values at very low temperatures) gives the following parameters for the global destruction rate coefficients of DME and MF through reaction with He$^+$: $\alpha=1.4 \times 10^{-9}$, $\beta$=$+0.295$, $\gamma$=$0.00$ for DME and $\alpha=1.4 \times 10^{-9}$, $\beta$=$+0.241$, $\gamma$=$0.00$ for MF. These values lead to rate coefficients much smaller than those adopted in astrochemical databases, especially at the low temperatures (below 50 K) of relevance in the ISM. The 
total rate coefficients must then be combined with the experimental BRs reported in Section \ref{sec:new-reaction-rates} to give the rates for the individual channels.

The product BRs were seen to be constant over a wide range of temperatures, so that we can reliably assume that they are also valid at the low temperatures of relevance to the ISM. Given that the channels with the largest yields, accounting globally for more than 80\% of the reactions, are those leading to HCO$^+$ and CH$_3^+$ for DME and the one leading to HCO$^+$ for MF, we recommend that only these selected channels are included when updating the astrochemical databases (the small yields of the other channels make the effects of their inclusion in astrochemical models negligible for all purposes).

Finally, we have to note that the present experimental results allow us to infer the chemical nature of the formed ionic products, but additional combinations of
neutral co-products are indeed possible. In particular:
\begin{enumerate}
\item {\it For DME}, as co-fragments of HCO$^+$, we can have either CH$_3$ + H$_2$, or
CH$_3$ + 2 H, while for CH$_3^+$ we can have either CH$_3$O or H$_2$CO + H;
 \item {\it For MF}, as co-fragments of HCO$^+$, we can have either CH$_3$O, CH$_2$OH, or HCO + H$_2$. 
\end{enumerate}
In the absence of any experimental information, we consider all of these combinations to be 
equally probable. We adopted a similar procedure to estimate the products of the DME
fragmentation from the reaction with  H$_3^+$ (see Section
\ref{sec:revis-valu-react}). Table \ref{tab:summary-rates} summarises the new values that we
recommend to be used in the astrochemical networks.

\begin{table*}
  \begin{tabular}{llcc}
    \hline
    
    Reactants & Products & $\alpha$ ($\times 10^{-10}$) & $\beta$ \\
    \hline
    CH$_3$OCH$_3$ + He$^+$ & He + CH$_3^+$ + CH$_3$O & 2.9 & 0.295 \\
    CH$_3$OCH$_3$ + He$^+$ & He + CH$_3^+$ + H$_2$CO + H & 2.9 & 0.295 \\
    CH$_3$OCH$_3$ + He$^+$ & He + HCO$^+$ + CH$_3$ + H$_2$ & 4.0 & 0.295\\ 
    CH$_3$OCH$_3$ + He$^+$ & He + HCO$^+$ + CH$_3$ + 2 H & 4.0 & 0.295\\ 
    HCOOCH$_3$ + He$^+$ & He + HCO$^+$ + CH$_3$O & 4.6  & 0.241 \\
    HCOOCH$_3$ + He$^+$ & He + HCO$^+$ + CH$_2$OH & 4.6 & 0.241 \\
    HCOOCH$_3$ + He$^+$ & He + HCO$^+$ + HCO + H$_2$ & 4.6 & 0.241 \\
    CH$_3$OCH$_3$ + HCO$^+$ & CO + CH$_3$OCH$_4^+$ & 21  & -0.50\\
    CH$_3$OCH$_3$ + H$_3^+$ & H$_2$ + CH$_3^+$ + CH$_3$O + H & 10 & -0.50\\
    CH$_3$OCH$_3$ + H$_3^+$ & H$_2$ + CH$_3^+$ + H$_2$CO + H$_2$ & 10 & -0.50\\
    CH$_3$OCH$_3$ + H$_3^+$ & H$_2$ + CH$_3$O$^+$ + CH$_4$ & 18 & -0.50\\
    CH$_3$OCH$_3$ + H$_3^+$ & H$_2$ + CH$_3$OHCH$_3$$^+$ & 8.4 & -0.50\\
    HCOOCH$_3$ + HCO$^+$ & CO + H$_5$C$_2$O$_2^+$ & 29 & -0.50\\
   HCOOCH$_3$ + H$_3^+$ & H$_2$ + CH$_3$OH$_2^+$ + CO & 34 & -0.50\\
   HCOOCH$_3$ + H$_3^+$ & H$_2$ + H$_5$C$_2$O$_2^+$ & 7.0 & -0.50\\
    \hline
  \end{tabular}
  \caption{Summary of products and rates of the reactions of He$^+$,
    HCO$^+$and H$_3^+$ with DME and MF, respectively. The parameter
    $\gamma$ is assumed to be equal to 0.00 on all reactions.
}
  \label{tab:summary-rates}
\end{table*}

\subsection{Impact on the abundances}

Our new rate coefficients for the destruction of DME and MF through reaction with He$^+$ differ greatly from those given by the astrochemical databases, KIDA and UDfA, especially at the low temperatures of the ISM gas. In particular, with respect to KIDA, our values are smaller by a factor of 5--6 at 100 K, the temperature of the warm gas in hot corinos
and outflow molecular shocks where iCOMs are abundantly detected (e.g. \citealt{bottinelli2004, bottinelli2007, Taquet2015, Jorgensen2016, Ligterink2017, Lefloch2012, Lefloch2017, Lefloch2018, Codella2017}), and by a factor of 25--28 at 10 K, the temperature of prestellar cores where iCOMs are also detected (\citealt{RN6, cernicharo2012, vastel2014, jimenez2016})

Additionally, the review of the literature regarding the DME and MF reactions with HCO$^+$ and H$_3^+$ has also shown important differences between the rate coefficients reported in the KIDA and UDfA databases with respect to the available experimental data, and more accurate experiments and/or theoretical calculations are clearly needed to better understand the products of the reactions and the dependence on temperature.

In this context, we ran a few simulations to quantify the effect of these differences on the predicted abundances. To this end, we used a version revised by us of the {\it Nahoon} code, 
which is publicly available in the KIDA database. Briefly, {\it Nahoon} is a pure gas-phase time-dependent code that allows to follow the chemical composition of a gas at a given temperature and density (\citealt{RN53}). In our case, we used $T$=10 K and $n_H$=$2\times10^5$ cm$^{-3}$ to describe a generic prestellar core, and $T$=100 K and $n_H$=$2\times10^8$ cm$^{-3}$ to describe a generic hot corino. 

Since the abundance of He$^+$ as well as HCO$^+$ and H$_3^+$ depends on the cosmic-ray ionisation rate, $\zeta_{CR}$, whose value in these regions is known with a large uncertainty (e.g. \citealt{padovani2009, vaupre2014}), we ran simulations using both a low ($10^{-18}$ s$^{-1}$) and high ($10^{-15}$ s$^{-1}$) cosmic ray ionisation rate to model two extreme situation of gas ionisation.

Finally, the initial abundances were chosen as follows. In the prestellar core models, we adopted the abundances from \cite{Jenkins2009}, with those of the heavy metals (S, Si, Mg, Fe, Na) being decreased by a factor of 100 and those for the light ones by a factor of ten: C/H=$1.7\times10^{-5}$, N/H=$6.2\times10^{-6}$, and we changed O/H ratio from $2.6\times10^{-5}$ to $1.8\times10^{-5}$ to simulate the segregation of oxygen on the water ices. In the hot corino and molecular outflow models, we used the initial abundances described in \cite{Codella2017}.
The parameters used in the models that we run are summarised in Table
\ref{tab:models}.
\begin{table*}
  \begin{tabular}{lcccc}
    \hline
    Model & Temperature & $n_H$     & $\zeta_{CR}$ & Initial abundances \\
              &  (K)                & cm$^{-3}$ & $10^{-17}$ s$^{-1}$       & \\
    \hline
    1        & 10        & $2\times 10^5$ & 0.1  & modified  Jenkins (2009); O/H=$2.6\times 10^{-5}$\\
    2        & 10        & $2\times 10^5$ & 0.1  & modified  Jenkins (2009); O/H=$1.8\times 10^{-5}$\\
    3        & 10        & $2\times 10^5$ & 100 & modified  Jenkins (2009); O/H=$2.6\times 10^{-5}$ \\
    4        & 10        & $2\times 10^5$ & 100 & modified  Jenkins (2009); O/H=$1.8\times 10^{-5}$\\
    5        & 100      & $2\times 10^8$ & 0.1  & Codella et al. (2017)\\
    6        & 100      & $2\times 10^8$ & 100 & Codella et al. (2017)\\
    \hline
  \end{tabular}
  \caption{Summary of the input values for the simulations that were run to study the impact of the new rate coefficients of the reactions of DME and MF with He$^+$, HCO$^+$ and H$_3^+$ on the predicted abundances of DME and MF in prestellar cores, and hot corinos and molecular outflow shocks, respectively. For the details on the initial abundances, see text.} \label{tab:models}
\end{table*}

We first run simulations with our old chemical network, which is based on
the KIDA 2014 network ({\it http://kida.obs.u-bordeaux1.fr}), upgraded
 with the reactions described in \cite{RN1, skouteris2018} and Russo et al. (2018). We then changed the reactions with He$^+$, HCO$^+$ and H$_3^+$ according to Table \ref{tab:summary-rates}
and studied the changes in the predicted abundances.
We note that our goal here is not to carry out a full study of the impact of the new reaction products and rate coefficients, but only to estimate whether this would dramatically change the model predictions in ``standard'' situations where DME and MF have been detected.

The effect is indeed dependent on the parameters that we have explored, details of which are given in Table \ref{tab:models}. In cold gas and conditions similar to the prestellar cores
the abundances of DME and MF can be different by up to 40\%, with the
difference larger when the oxygen abundance is greater (models 1 and
3). This effect does not seem to depend much on the cosmic-ray ionisation rate used. This is also the case for the models simulating the hot corino gas, where the difference in the abundances are smaller (about $\leq 10$\%), as expected based on the difference in the rate coefficients at 100 K.
This is because, in all these models, the DME and MF destruction is
dominated by H$_3^+$, with the HCO$^+$ and He$^+$ pathways proceeding at a slower rate by about a factor of ten.
We emphasize that, although our pure gas-phase model predicts low ($\leq 10^{-14}$) abundances for DME and MF, the effect may be larger if a full model, such as that used in \cite{RN1}, is employed, though this is beyond the scope of this work.
%We emphasize, thought, that our pure gas-phase model predicts low ($\leq 10^{-14}$) abundances for DME and MF so that the effect may be larger if a full model, like in \cite{RN1} is taken into account. This is out of the scope of this work though.

%===============================================
\section{Conclusions}\label{sec:conclusions}

We have reported a new study of the branching ratios, products and rate coefficients for the reactions of DME and MF with He$^+$. 
We have also reviewed the destruction routes of DME and MF by the other two abundant ions in molecular gas, namely H$_3^+$ and HCO$^+$.
The main conclusions of this work are:
\begin{itemize}
\item We have computed branching ratios, products and rate coefficients of DME/MF + He$^+$. These calculations are based on the recent measurements and theoretical modelling of the cross-sections performed by \citet{C7CP00827A, cernuto2018}. 
\item The new DME/MF + He$^+$ branching ratios,
  products and rate coefficients substantially differ from those
  reported in the commonly used astrochemical databases KIDA and UDfA.
\item The dependence of the rate coefficients on the temperature is
 positive and not negative, as assumed in the KIDA and UDfA databases. As a consequence, the difference in the rate coefficients is larger at low temperatures: at 10 K the values differ by more than a factor of 20. 
\item The review of the DME/MF + H$_3^+$/HCO$^+$ reactions show that the BRs, products and rate coefficients of these reactions are poorly understood. Nonetheless, the data available in the literature
  differ from that used in the KIDA and UDfA databases.
\item We provide a table of the parameters to include in the
  astrochemical networks that we recommend for the six
  reactions of DME and MF with He$^+$, H$_3^+$ and HCO$^+$. 
\item A very general modelling of cold (10 K) and warm (100 K) gas
  simulating the conditions were DME and MF are detected, namely
  prestellar cores and hot cores/corinos, shows that the new rates have
  an impact on the predicted abundances, of up to $\sim40\%$. However, different conditions might lead to even larger differences.
\end{itemize}
In conclusions, we emphasise the importance to have reliable
information on the destruction pathways of iCOMs by the most  abundant ions in molecular gas: H$_3^+$, HCO$^+$ and He$^+$. This can be obtained with experimental works, as in the case reported here 
\citep{C7CP00827A, cernuto2018} coupled with theoretical modelling, or by ab initio quantum chemistry calculations.

A more general conclusion is that attention should be paid when using estimated rate coefficients for unknown processes or when extrapolating room temperature rate coefficients at very low temperatures. In the case of the reactions between  He$^+$ and DME/MF the results obtained with a very detailed treatment of the entrance channel challenge the common notion that ion-molecule reaction rate coefficients increase with T (or remain constant as in the Langevin model). In addition, we note that product branching ratios are difficult to guess in the absence of experimental or theoretical results. We recommend that the physical chemistry community dedicate as much effort as possible to characterize all the reactions which play a significant role in astrochemical models.

%===============================================
\begin{acknowledgements}

  We thank Mr. Vincent Richardson for his precious help with English Language editing and revision
  This work was supported by the the Italian Ministero
  dell'Istruzione, Universit\'a e Ricerca through the grant Progetti
  Premiali 2012 - iALMA (CUP C52I13000140001).
  This project has received funding from the European Research Council
  (ERC) under the European Union's Horizon 2020 research and
  innovation programme, for the Project “The Dawn of Organic
  Chemistry” (DOC), grant agreement No 741002.
  The support of the Department of Physics of the University of Trento is gratefully acknowledged.
  N.B. and F.P. acknowledge the financial support from Italian MIUR "PRIN 2015", project "STARS in the CAOS (Simulation Tools for Astrochemical Reactivity and Spectroscopy in the Cyberinfrastructure for Astrochemical Organic Species)," grant number 2015F59J3R.

\end{acknowledgements}

%===============================================
\bibliographystyle{aa}
\bibliography{biblio}
%===============================================

\end{document}